\documentclass[aip,jcp,reprint,author-year,author-numerical]{revtex4-1}
\usepackage{amsmath}
\usepackage{amsfonts}
\usepackage{amssymb}
\usepackage{bm}% bold math
\usepackage[utf8]{inputenc}
\usepackage[T1]{fontenc}
\usepackage{newtxmath}
\usepackage{etoolbox}
\makeatletter
\def\@email#1#2{%
 \endgroup
 \patchcmd{\titleblock@produce}
  {\frontmatter@RRAPformat}
  {\frontmatter@RRAPformat{\produce@RRAP{*#1\href{mailto:#2}{#2}}}\frontmatter@RRAPformat}
  {}{}
}%
\makeatother
\usepackage{xcolor}
\usepackage{makeidx}
\usepackage{graphicx}
\usepackage{tikz}
\usepackage{float}
\usepackage{braket}
\usepackage{lipsum}
\usepackage{enumitem}
\usepackage[caption=false]{subfig}
\usepackage{color}
\usepackage[colorlinks=true,citecolor=blue,linkcolor=blue,urlcolor=blue]{hyperref}
\usepackage{url}
\usepackage{braket}
\usepackage{verbatim}

\newcommand{\beq}{\begin{equation}} 							
\newcommand{\eeq}{\end{equation}}
\newcommand{\bematrix}{\left(\begin{matrix}}
\newcommand{\ematrix}{\end{matrix}\right)}
\newcommand{\id}{\mathcal{I}}

\begin{document}

\title{Polaron Transformed Canonically Consistent Quantum Master Equation}

\author{Juzar Thingna}
\affiliation{American Physical Society, 100 Motor Parkway, Hauppauge, New York 11788, USA}
\affiliation{Center for Theoretical Physics of Complex Systems, Institute for Basic Science (IBS), Daejeon 34126, Republic of Korea}

\author{Xiansong Xu}
\affiliation{College of Physics and Electronic Engineering and Center for Computational Sciences, Sichuan Normal University, Chengdu 610068, China}

\author{Daniel Manzano}
\affiliation{Quantum Thermodynamics and Computation Group. Departamento de Electromagnetismo y Física de la Materia, Universidad de Granada, 18071 Granada, Spain}
\affiliation{Instituto Carlos I de Física Teórica y Computacional, Universidad de Granada, 18071 Granada, Spain.}

\date{\today}
\begin{abstract}
A central challenge in the theory of open quantum systems is the development of theoretical frameworks capable of accurately describing large, strongly interacting quantum many-body systems in the regime of strong system–bath interaction. In this work, we take a step toward this goal by formulating a polaron-transformed version of the canonically consistent quantum master equation (CCQME) [T. Becker \emph{et~al.}, Phys. Rev. Lett. \textbf{129}, 200403 (2022)]. The CCQME extends beyond standard weak-coupling approaches while retaining the same numerical complexity as conventional quantum master equations, thereby enabling the treatment of large quantum systems. The polaron transformation further enhances the accessible system–bath interaction strengths, allowing us to move from moderate to ultra-strong interaction regimes. We present a unified and transparent derivation of these two approaches and combine them to obtain the polaron-transformed CCQME (PT-CCQME). Applying our method to the paradigmatic spin–boson model, we find excellent agreement with numerically exact time-evolving matrix product operator (TEMPO) simulations. Finally, we predict an initial-state-independent \emph{slowing} down of thermalization in the strong-coupling regime of the spin-boson model.
\end{abstract}

\maketitle

\section{Introduction}
The study of open quantum systems (OQS) has traditionally relied on the assumption of weak interaction between the system and its environment. This premise is the cornerstone of Markovian master equations, most notably the Redfield and Lindblad formalisms \cite{Redfield1965,Davies1974, Lindblad1976, Breuer2003, Manzano2020}. However, a vast array of prominent quantum systems—including quantum dots \cite{Vora2015, Crooker2002}, photosynthetic complexes \cite{Brixner2005, Manzano2013, Yang2020}, complex networks \cite{thingna16,manzano2016,Manzano2014},  and various chemical reaction centers \cite{YangCao2021, Lindoy2023}—operate in regimes of strong system-bath interaction that defy this approximation. Developing accurate master equations for this strong-coupling regime presents significant challenges, ranging from long-time inaccuracies \cite{Mori08, Fleming2011, Thingna12, Thingna13} and high computational overhead to the emergence of unphysical non-positive states when perturbative expansions are truncated at the second order \cite{Hartmann2020}. While rigorous frameworks such as higher-order expansions \cite{Jang02, Thingna14, Crowder24}, reaction-coordinate mapping~\cite{Nazir18, AntoSztrikacs23}, the density matrix renormalization group (DMRG) \cite{Schollwoeck2005, Prior2010}, the multi-configuration time-dependent Hartree (MCTDH) method \cite{Beck2000, Wang2003}, and the hierarchical equations of motion (HEOM) \cite{Tanimura2020} have been developed to address these issues, they often incur a prohibitive computational cost.

Investigating the strong-coupling regime is not merely a computational challenge but a fundamental necessity for understanding quantum dynamics in realistic environments. As the interaction strength increases, the clear separation of time scales between the system and the bath—a prerequisite for the Markovian approximation—frequently breaks down. This leads to pronounced non-Markovian effects, characterized by a backflow of information from the bath to the system and memory-dependent dynamical evolution \cite{Breuer2016, deVega2017, Hartmann25}. Capturing these non-Markovian features is critical for modeling biological light-harvesting, where the interplay between coherent dynamics and environmental noise is thought to enhance transport efficiency \cite{ThingnaPRB12, Chin2013, Zhou15}, as well as for the precise control of solid-state qubits, where environmental memory can significantly alter decoherence pathways \cite{Kloeffel2013}.

In this work, we address these challenges by combining two promising approaches: the Canonically Consistent Quantum Master Equation (CCQME) and the polaron transformation. The CCQME \cite{Becker2022, Sartipi26} is a fourth-order perturbative technique that provides a thermodynamically consistent framework, ensuring the system relaxes to the correct mean-force Gibbs state regardless of the interaction strength. This effectively resolves the unphysical steady-state errors common in standard weak-coupling master equations. Furthermore, the CCQME offers a significant practical advantage: it avoids the cumbersome super-operators involving multidimensional integrals and the elaborate numerical resources required by exact methods. In Ref.~[\onlinecite{Becker2022}], this method was benchmarked against standard Redfield and Lindblad equations in solvable models, demonstrating superior precision and stability.

Complementing the CCQME is the polaron transformation \cite{Jang11,Lee12,xu16,Hsied19}, a technique based on ``dressing'' the quantum system with bath modes. This unitary transformation shifts the problem to a frame where the dressed system (the polaron) incorporates the dominant effects of the system-bath interaction. Consequently, the remaining interaction term in the new frame is often sufficiently weak to be treated perturbatively. This allows for an accurate description of strong-coupling dynamics and incoherent hopping using standard master equation tools, which would otherwise fail in the original frame.

However, while the polaron master equation is a powerful tool, it is not without limitations. Standard derivations often rely on second-order Born-Markov approximations applied to the residual interaction, which can lead to inaccuracies in intermediate coupling regimes or at high temperatures where the residual term is not negligible \cite{Jang2008, McCutcheon_2010}. By applying the CCQME formalism within the polaron frame, we aim to treat these residual interactions with fourth-order accuracy while explicitly maintaining thermodynamic consistency. This hybrid approach allows us to bridge the gap between perturbative efficiency and non-perturbative accuracy, offering a robust tool for describing quantum dynamics in regimes where neither method suffices in isolation.

This paper is organized as follows. In Sec. \ref{sec:theory}, we derive the general framework to apply the polaron transformation to the CCQME. This framework is applied to the specific case of a spin-boson model in Sec.~\ref{sec:sbm} and numerically analyzed in Sec. \ref{sec:numerics}. Finally, we end with some concluding remarks in Sec.~\ref{sec:discussion}. 

\section{Theoretical Framework}
\label{sec:theory}
In this section, we derive the polaron-transformed, canonically consistent quantum master equation in a general, system-independent manner.

\subsection{Model}
We consider a standard open quantum system model consisting of a system (described by a Hamiltonian $H_S$) connected to a bath ($H_B$) via a system-bath interaction ($H_{SB}$). The total Hamiltonian~\footnote{The above Hamiltonian does not include the renormalization term (also commonly known as the counter term) of the Caldeira-Leggett model, however, our approach outlined in this work can be easily adapted to include this term if needed.} is given by
\begin{equation}
    H = H_S + H_B + H_{SB},
    \label{eq_total}
\end{equation}
and the state of the system is given by a density matrix
$\rho \in\mathcal{H}_S \otimes \mathcal{H}_B$. The bath is modeled as an infinite collection of free bosons with frequencies $\omega_k$, described by the Hamiltonian
\begin{equation}
    H_B = \sum_{k}\omega_k b_k^\dagger b_k,
    \label{eq_bath}
\end{equation}
where $b_k$ ($b_k^\dagger$) are the bosonic annihilation (creation) operators, respectively. The system-bath interaction is assumed to be of the form 
\begin{equation}
    H_{SB} = \sum_{n,k} g_{nk} |n\rangle \langle n|\left(b_k^{\dagger} + b_k\right),
    \label{eq_systembath}
\end{equation}
where $\{|n\rangle\}$ are the eigenstates of the system-bath interaction operator. The bath interaction term, proportional to $(b_k^\dagger + b_k)$, corresponds to the bath position operator.

In the basis defined by the states $|n\rangle$, the most general form of the system Hamiltonian is
\begin{equation}
    H_S = \sum_n \varepsilon_n |n\rangle \langle n| + \sum_{m\neq n} h_{mn} |m\rangle \langle n|
    \label{eq:system}
\end{equation}
where $\varepsilon_n \in \mathbb{R}$ and $h_{mn} = h_{nm}^* \in \mathbb{C}$. Although this Hamiltonian is expressed in a specific basis, we impose no constraints on the range or strength of the interactions, rendering it a general description of the system. The only key assumptions are that the bath consists of free bosons and that the system couples linearly to the bath's position coordinate. This standard assumption in open quantum system theory ensures that the bath acts as a proper thermal reservoir \cite{Chen1989,Rivas2010,Garrido2024,Asadian2013}.

\subsection{Polaron Transformation}
\label{sec:polaron}

The polaron technique~\cite{Rackovsky1973,Abram1975,Silbey1984,Jang2008,Jang11,Lee12,xu16,Hsied19} is an exact canonical transformation of the full Hamiltonian that renders an originally strong system-bath interaction to a weak one in a new frame. The transformation is governed by the operator
\begin{align}
\label{eq:transform}
T &= \sum_{n,k} \frac{g_{nk}}{\omega_k} |n\rangle \langle n| \otimes (b_k^\dagger - b_k),\nonumber \\
e^{T}& = \sum_n |n\rangle \langle n| \otimes \Pi_{k} D_{nk}^{\dagger},
\end{align}
where $D_{nk} = \exp[g_{nk}(b_k - b_k^\dagger)/\omega_k]$ is the displacement operator for the $k$-th mode acting on the system subspace $\ket{n}\bra{n}$ and $\Pi_k$ is used to denote the product of a sequence of terms such that $\Pi_k a_k = a_1\cdot a_2, \cdots a_\infty$ for an arbitrary operator $a_k$. While we provide a brief overview of the polaron technique in this subsection for completeness, we direct interested readers to Ref.~[\onlinecite{Jang11}] for an illuminating, in-depth discussion.

This operator is used to transform the state and Hamiltonian to the polaron frame. In this frame, the bath Hamiltonian [Eq.~\eqref{eq_bath}] transforms as,
\begin{equation}
e^{T} H_B e^{-T} = \sum_n |n\rangle \langle n| \sum_k \omega_k D_{nk}^\dagger b_k^\dagger b_k D_{nk}.
\end{equation}
Above we have used the definition of the polaron transform in terms of the displacement operator and the orthogonality of the states $\ket{n}$. Using the Baker-Campbell-Hausdorff relation ($e^X Y e^{-X} = Y + [X,Y] + [X,[X,Y]] + \cdots$) for the annihilation operator $b_k$ we obtain
\begin{equation}
    D_{nk}^\dagger b_k D_{nk} = b_k - \frac{g_{nk}}{\omega_k}.
\end{equation}
Using the above relation, we obtain 
\begin{align}
    e^{T} H_B e^{-T} &= \sum_k \omega_k b_k^\dagger b_k - \sum_{n,k} g_{nk}|n\rangle \langle n| \left(b_k^\dagger + b_k\right)\nonumber \\
 & + \sum_{n,k} \frac{g_{nk}^2}{\omega_k} |n\rangle \langle n|.
\end{align}
Above and throughout the text we have assumed $g_{nk} \in \mathbb{R}$. Similarly, the polaron transformed system-bath interaction via Eq.~\eqref{eq_systembath} reads,
\begin{align}
e^{T}H_{SB}e^{-T}&=\sum_{n,k} g_{nk} |n\rangle \langle n|\left(b_k^{\dagger} + b_k\right)- 2\sum_{n,k} \frac{g_{nk}^2}{\omega_k} |n\rangle \langle n|.
\end{align}
Lastly, the transformed system Hamiltonian [Eq.~\eqref{eq:system}] reads,
\begin{equation}
    e^{T} H_S e^{-T} = \sum_n \varepsilon_n |n\rangle \langle n| + \sum_{m\neq n} h_{mn} |m\rangle \langle n| \otimes \Theta_m^\dagger \Theta_n
\end{equation}
with $\Theta_n=\Pi_k D_{nk}$. Thus, the total Hamiltonian in the polaron frame reads,
\begin{align}
   e^{T} H e^{-T} &= \sum_n \left(\varepsilon_n-\sum_{k} \frac{g_{nk}^2}{\omega_k}\right) |n\rangle \langle n| + \sum_k \omega_k b_k^\dagger b_k \nonumber \\  
    &+ \sum_{m\neq n} h_{mn} |m\rangle \langle n| \otimes \Theta_m^\dagger \Theta_n.
\end{align}

Next we center the bath such that the average interaction energy between the system and bath is zero, i.e, adding and subtracting $\sum_{m\neq n} h_{mn} |m\rangle \langle n| \langle \Theta_m^\dagger \Theta_n\rangle$ we obtain,
\begin{equation}
e^{T} H e^{-T} \equiv \tilde{H} = \tilde{H}_S + \tilde{H}_B + \tilde{H}_{SB}
\label{eq:htilde}
\end{equation}
with the effective bath Hamiltonian remaining unchanged
\begin{equation}
    \tilde{H}_B = H_B = \sum_k \omega_k b_k^\dagger b_k.
\end{equation}
The \emph{residual} system-bath interaction takes the form
\begin{align}
\tilde{H}_{SB}=\sum_{n \neq m} h_{mn} |m\rangle \langle n| V_{mn} 
\equiv \sum_{n \neq m} S_{mn} \otimes V_{mn}
\label{eq:systembathint}
\end{align}
with $S_{mn} = h_{mn} |m\rangle \langle n|$ and
\begin{equation}
\label{eq:V}
V_{mn}=\Theta^\dagger_m\Theta_n - \kappa_{mn}.
\end{equation}
Moreover, the effective system Hamiltonian reads,
\begin{equation}
\label{eq:polaronH}
\tilde{H}_S = \sum_n \left(\varepsilon_n-\sum_{k} \frac{g_{nk}^2}{\omega_k}\right) |n\rangle\langle n| + \sum_{m \neq n} \kappa_{mn} h_{mn} |m\rangle\langle n|.
\end{equation}
In the special case when $|g_{nk}| = |g_k|$, the onsite energies $\varepsilon_n$ are affected only by a constant shift $-\sum_k g_k^2/\omega_k$ for all $n$, making the renormalization irrelevant. The renormalization factor $\kappa_{mn}$, which ensures $\langle V_{mn}\rangle =0$, due to the centering of the bath, is given by
\begin{equation}
\kappa_{mn} = \langle \Theta_m^\dagger \Theta_n \rangle = \exp\left[-\frac{1}{2} \sum_k \frac{\delta g_{mn,k}^2}{\omega_{k}^2} \coth\left(\frac{\beta \omega_{k}}{2}\right)\right].
\label{eq:kappa}
\end{equation}
Above $\langle \cdots \rangle = \mathrm{Tr}_B [\cdots \exp[-\beta \tilde{H}_B]/\tilde{Z}_B] $ with $\tilde{Z}_B=\mathrm{Tr}_B [\exp[-\beta \tilde{H}_B]/\tilde{Z}_B] $ being the bath partition function, $\delta g_{mn,k} = g_{mk}-g_{nk}$, and $\beta$ is the inverse temperature of the bath. Throughout this work we consider $k_B= \hbar =1$.

\subsection{Canonically Consistent Quantum Master Equation in the Polaron Frame}
\label{sec:PTCCQME}
The Canonically Consistent Quantum Master Equation (CCQME), recently introduced in Ref.~[\onlinecite{Becker2022}], provides a framework for describing open quantum system dynamics beyond the traditional weak-coupling limits of Redfield~\cite{Redfield1965} and Lindblad~\cite{Lindblad1976} theories. By incorporating the mean-force Gibbs state~\cite{Trushechkin22} from equilibrium statistical mechanics, the CCQME ensures that the long-time steady state remains consistent with the Hamiltonian of mean force up to second order in the system-bath interaction strength~\cite{Thingna12}. Benchmarks against exactly solvable models have demonstrated that the CCQME significantly improves accuracy in strong-coupling and low-temperature regimes—correctly reproducing ground-state populations and mitigating positivity violations—making it a versatile tool for quantum thermodynamics, many-body physics, and quantum transport.

To derive the Polaron-Transformed Canonically Consistent Quantum Master Equation (PT-CCQME), we begin with the evolution of the polaron transformed total density matrix $\tilde{\rho}(t) = e^{T}\rho(t) e^{-T}$ that follows the von Neumann equation
\begin{equation}
    \frac{d\tilde{\rho}(t)}{dt}= -i [\tilde{H}, \tilde{\rho}(t)],
\end{equation}
where $\tilde{H}=e^{T} H e^{-T}$ is the polaron transformed total Hamiltonian given in Eq. \eqref{eq:htilde}.

Following the procedure in Ref.~[\onlinecite{Becker2022}], we evaluate the Dyson map in the polaron frame, $\tilde{\Lambda}_t$, for the reduced density matrix $\tilde{\rho}_S(t) = \tilde{\Lambda}_t[\tilde{\rho}_S(0)]$. To second order in the residual system-bath interaction this map is given by~\cite{Becker2022,Breuer2003} 
\begin{equation}
\tilde{\Lambda}_t[\bullet] \simeq \tilde{\Lambda}_t^0 \left[ \id[\bullet] + \int_0^t d\tau \tilde{\Lambda}_{-\tau}^0 [ \tilde{\mathcal{R}}_\tau [\tilde{\Lambda}_\tau^0 [\bullet] ] ] \right],
\label{eq:dyson}
\end{equation}
where $\tilde{\Lambda}_t^0 [\bullet]=e^{-i \tilde{H}_St} \bullet e^{i\tilde{H}_St}$, $\id[\bullet] = \bullet$ is the identity super-operator, and $\tilde{\mathcal{R}}_t[\bullet]$ is the time-dependent Redfield dissipator. We assume decoupled initial condition in the polaron frame $\tilde{\rho}(0)=\tilde{\rho}_S(0)\otimes \rho_B(0)$, which is consistent with a decoupled state in the original frame provided $\left[ T,\rho_S(0) \right]=0$\footnote{The method can be generalized for cases $\left[ T,\rho_S(0) \right]\neq0$, however, in this case the Dyson map in the polaron frame will also have an initial value terms similar to the ones described in Refs.~[\onlinecite{Jang11}, \onlinecite{ThingnaPRB14}, \onlinecite{Chang12}]}. Throughout this work we will assume the bath in the original frame to be in an initial thermal state, i.e., $\rho_B(0)=\exp[-\beta H_B]/Z_B$ with $Z_B = \mathrm{Tr}_B[\exp[-\beta H_B]]$ being the partition function of the bath. Note that as $H_B=\tilde{H}_B$, the partition function of the bath is the same in both the original and polaron frames, $Z_B=\tilde{Z}_B$.

To avoid the linear temporal divergences inherent in truncated Dyson expansions~\cite{Thingna14}, we transition to a differential equation representation for the reduced density matrix that contains no divergences. Thus, one obtains a Redfield-like inhomogeneous non-Markovian equation
\begin{equation}
    \frac{d\tilde{\rho}_S(t)}{dt} = -i [\tilde{H}_S, \tilde{\rho}_S(t)] + \tilde{\mathcal{R}}_t [ \tilde{\Lambda}_t^0 [\tilde{\rho}_S(0)] ].
    \label{eq:redfield}
\end{equation}
The Redfield dissipator is defined as,
\begin{align}
\label{eq:Redfield}
\tilde{\mathcal{R}}_t[\bullet] = \sum_{\substack{n \neq m \\ n' \neq m'}} \int_{0}^{t} & d\tau \, \left\{ [ S_{m'n'}(-\tau)\bullet, S_{mn} ] C_{m'n'}^{mn}(\tau) \right.\nonumber \\
& \left.- [ \bullet \, S_{m'n'}(-\tau), S_{mn}] C_{mn}^{m'n'}(-\tau)\right\},
\end{align}
where $S_{mn}(-\tau) = e^{-i\tilde{H}_S \tau} S_{mn} e^{i \tilde{H}_S \tau}$ is the interaction picture representation of the operator $S_{mn}$. Under the Born approximation~\cite{Breuer07}, i.e, $\tilde{\rho}_B(\tau) \approx \tilde{\rho}_B(0) = \rho_B(0)$, the bath-bath correlation functions $C_{m'n'}^{mn}(\tau) = \mathrm{Tr}_B[V_{mn}(\tau) V_{m'n'} \rho_B(0)]$ are expressed as (see Appendix~\ref{sec:corr2})
\begin{equation}
\label{eq:C}
C^{mn}_{m'n'}(\tau) = \kappa_{mn}\kappa_{m'n'}\left[e^{-\epsilon^{m'n'}_{mn}}-1\right],
\end{equation}
with the exponent defined by the spectral properties of the environment:
\begin{align}
\label{eq:epsilon}
    \epsilon_{m'n'}^{mn} = \sum_k \frac{\delta g_{mn,k}\delta g_{m'n',k}}{\omega_k^2}&\left[\coth{\left(\frac{\beta\omega_k}{2}\right)}\cos\left(\omega_k\tau\right)\right. \nonumber \\
    &\left.-i\sin\left(\omega_k\tau\right)\right].
\end{align}%

Markovian substitution of Eq.~\eqref{eq:redfield} via the approximation $\tilde{\Lambda}_t^0[\tilde{\rho}_S(0)] \simeq \tilde{\rho}_S(t)$, which is consistent with the weak-coupling approximation, and replacing $\tilde{\mathcal{R}}_t$ with $\tilde{\mathcal{R}}_\infty$ yields the polaron transformed Redfield equation (PT-Redfield). The CCQME is based on refining this part of the weak-coupling approximation by defining a second-order residual super-operator $\tilde{Q}_t$ such that
\begin{equation}
\tilde{\rho}_S(t)\simeq \left( \id +  \tilde{Q}_t \right)[ \tilde{\Lambda}_t^0 \tilde{\rho}_S(0)].
\label{eq:lambda_q}    
\end{equation}
The intuition behind this form comes from the Dyson map [see Eq.~\eqref{eq:dyson}] which, if acted upon $\tilde{\rho}_S(0)$ gives the above form. Inverting the above relation using $\left( \id + \tilde{Q}_t \right)^{-1} \simeq \left( \id - \tilde{Q}_t \right)$ leads to the master equation 
\begin{equation}
\label{eq:CCQMEQt}
\frac{d\tilde{\rho}_S(t)}{dt} = -i [\tilde{H}_S, \tilde{\rho}_S(t)] + \tilde{\mathcal{R}}_t \left[ (\id - \tilde{Q}_t)[\tilde{\rho}_S(t)] \right].   
\end{equation}
However, since the $\tilde{Q}_t$ obtained from the Dyson map diverges linearly in time, the above equation is unstable. Thus, to ensure stability and physical consistency, the CCQME replaces the potentially divergent $\tilde{Q}_t$ with its equilibrium analogue, the mean-force Gibbs (MFG) correction $\tilde{Q}^{\mathrm{MFG}}$.

In the ultra-strong-coupling limit, the mean-force Gibbs state is a closed-form expression of the equilibrium state of the system~\cite{CresserPRL21,Trushechkin22}. The MFG state can be understood as a state obtained by weakly coupling the total Hamiltonian to a super-bath with same temperature $\beta$ and then tracing out the bath degrees of freedom. Thus up to second order in residual interaction the MFG is given by,
\begin{equation}
\lim_{t\to\infty} \tilde{\rho}_S(t) = \frac{\mathrm{Tr}_{B}[e^{-\beta \tilde{H}}]}{\tilde{Z}} \simeq \left( \id +  \tilde{Q}^{\mathrm{MFG}} \right) \left[\frac{e^{-\beta \tilde{H}_S}}{\tilde{Z}_S}\right],
\label{eq:MFG}
\end{equation}
where $\tilde{Z}=\mathrm{Tr}[e^{-\beta \tilde{H}}]$ and $\tilde{Z}_S=\mathrm{Tr}_{S}[e^{-\beta \tilde{H}_S}]$ are the partition functions of the total system and system, respectively~\footnote{Note that since the inhomegenous term in Eq.~\eqref{eq:redfield} is in the polaron frame, we also require the mean-force Gibbs correction $\tilde{Q}^{\mathrm{MFG}}$ to be in the polaron frame. Moreover, it is easy to see that the MFG from the original frame to the polaron frame can be obtained by replacing $H$ with $\tilde{H}$.}. 

Taking the standard Markovian limit in Eq.~\eqref{eq:CCQMEQt}, which assumes that the bath correlation functions decay rapidly, allows us to replace the time-dependent Redfield dissipator $\tilde{\mathcal{R}}_t$ and the transient Dyson map super-operator $\tilde{Q}_t$ with their asymptotic counterparts, $\tilde{\mathcal{R}}_\infty$ and $\tilde{Q}_\infty$. At this stage, we introduce a physically motivated substitution by replacing the dynamical super-operator $\tilde{Q}_\infty$ with the statistical mean-force Gibbs correction $\tilde{Q}^{\textrm{MFG}}$ defined in Eq.~\eqref{eq:MFG}. In the long-time limit, these two super-operators are expected to coincide despite their distinct origins, reflecting the fundamental requirement that an open quantum system must thermalize to the equilibrium state predicted by statistical mechanics. However, explicitly demonstrating this equivalence at the perturbative level is obscured by the fact that $\tilde{Q}_\infty$ diverges. This secular divergence is purely an artifact of perturbative truncation; the full, untruncated Dyson map remains entirely divergence-free. Consequently, there is a strict non-commutativity between taking the long-time limit $t\rightarrow \infty$ and truncating the Dyson series to a finite order (see Ref.~[\onlinecite{Thingna14}] for a detailed discussion of such divergences in higher-order quantum master equations). By substituting $\tilde{Q}_\infty$ with $\tilde{Q}^{\textrm{MFG}}$, we effectively bypass this non-commutativity, ensuring that the infinite-time limit is taken prior to the second-order truncation. This regularization ultimately yields the time-local PT-CCQME: 
\begin{equation}
\label{eq:CCQME}
\frac{d\tilde{\rho}_S(t)}{dt} = -i [\tilde{H}_S, \tilde{\rho}_S(t)] + \tilde{\mathcal{R}}_\infty \left[ (\id - \tilde{Q}^{\mathrm{MFG}})[\tilde{\rho}_S(t)] \right].   
\end{equation}
We stress here that the PT-CCQME is not exact, though it successfully captures significant physical effects up to fourth order in the residual interaction strength. Furthermore, its derivation relies on the standard Born-Markov approximation~\cite{Breuer07} alongside the assumption of a completely decoupled initial system-bath state. Specifically, the Born approximation assumes that the bath is sufficiently large such that its state remains time-independent and effectively immune to any backaction from the smaller system. Concurrently, the Markov approximation assumes that the system's dynamics are memoryless; its evolution at any given time $t$ depends solely on its state at $t-\delta t$, completely discarding any dependence on its prior dynamical history. By construction, this formalism utilizes equilibrium statistical information to systematically correct the quantum dynamics, ensuring that the steady state matches the MFG state up to second order in the residual interaction (see Appendix~\ref{sec:WhyPolaron}).

In Ref.~[\onlinecite{Becker2022}], the authors derived the CCQME for a specific form of the system-bath interaction $H_{SB}= S\otimes B$, however, since the system-bath interaction in the polaron frame is of a more general form [see Eq.~\eqref{eq:systembathint}] we re-derive the expression for $\tilde{Q}^{\mathrm{MFG}}$ in a more general way (see Appendix~\ref{sec:Q}), yielding
\begin{equation}
\label{eq:Qfinal}
\tilde{Q}^{\mathrm{MFG}}[\tilde{\rho}_S(t)] = \mathbf{P}\frac{1}{i\tilde{\Delta}}[\tilde{\mathcal{R}}_{\infty}[\tilde{\rho}_S(t)]] + \tilde{\mathcal{L}}[\mathbf{P^c}\,\tilde{\rho}_S(t)].
\end{equation}
Here $\mathbf{P} = \sum_{n\neq m}\mathbf{P}_{nm}$ is the projector into the coherent subspace of $\tilde{H}_S$ with $\mathbf{P}_{nm}\bullet = |\tilde{n}\rangle\langle \tilde{n}| \bullet |\tilde{m}\rangle\langle \tilde{m}|$, $|\tilde{n}\rangle$ is an eigenstate of $\tilde{H}_S$ with eigenenergy $E_n$, and $\mathbf{P^c}=\sum_n \mathbf{P}_{nn}$ projects into the complementary subspace to $\mathbf{P}$. The super-operator $(i\tilde{\Delta})^{-1}[\bullet]$ is defined such that
\begin{equation}
    \mathbf{P} \frac{1}{i\tilde{\Delta}}[\bullet] := \sum_{n \neq m} \frac{1}{i\delta E_{nm}}|\tilde{n}\rangle\langle \tilde{n}| \bullet |\tilde{m}\rangle\langle \tilde{m}|,
\end{equation}
where $\delta E_{nm} = E_n - E_m$. Equation~\eqref{eq:Qfinal} becomes singular in the presence of degeneracies (i.e., when $\delta E_{nm}=0$); therefore, in this work, we restrict our analysis to the non-degenerate case. Above, $\tilde{\mathcal{R}}_\infty$ is the Redfield dissipator given in Eq.~\eqref{eq:Redfield} and the generalized Lindblad~\footnote{In the standard Lindblad form $\Gamma_{uj} = \Gamma_{uj}'$.} dissipator $\tilde{\mathcal{L}}$ takes the form,
\begin{equation}
\label{eq:Qdiag}
\tilde{\mathcal{L}}[\bullet] = \sum_{u,j} \left( \Gamma_{uj}L_{uj} \bullet L_{uj}^{\dagger} - \frac{\Gamma_{uj}'}{2}\{ L_{uj}^\dagger L_{uj},\bullet \}\right) ,
\end{equation}
with $L_{uj}=|\tilde{u}\rangle\langle \tilde{j}|$ being the jump operators and the generalized rates 
\begin{align}
\label{eq:Lindbladrates}
\Gamma_{uj} &= \sum_{\substack{n \neq m \\ n' \neq m'}} \langle \tilde{u}|S_{mn}|\tilde{j}\rangle\langle \tilde{j} |S_{m'n'}|\tilde{u}\rangle \mathrm{Im}\left[\frac{\partial W^{m'n'}_{mn}(\delta E_{uj})}{\partial \delta E_{uj}}\right], \nonumber \\
\Gamma_{uj}' &= \sum_{\substack{n \neq m \\ n' \neq m'}} \langle \tilde{u}|S_{m'n'}|\tilde{j}\rangle\langle \tilde{j} |S_{mn}|\tilde{u}\rangle \mathrm{Im}\left[\frac{\partial W^{mn}_{m'n'}(\delta E_{uj})}{\partial \delta E_{uj}} \right. \nonumber\\
&\left.\qquad\qquad\qquad\qquad\qquad\qquad +\beta W^{mn}_{m'n'}(\delta E_{uj})\right].
\end{align}
Where we have defined the functions $W^{mn}_{m'n'}(\lambda) = \int_0^{\infty} d\tau \, C^{mn}_{m'n'}(\tau) e^{-i\lambda\tau}$, and $C^{mn}_{m'n'}(\tau)$ are the correlation functions given by Eqs.~\eqref{eq:C}--\eqref{eq:epsilon}. The functions $W^{mn}_{m'n'}(\lambda)$ also appear in the Redfield equation as \emph{rates}~\footnote{Only the real part of $W^{mn}_{m'n'}(\delta)$ is referred to as rate in the Redfield equation, whereas the imaginary part is known as Lamb shifts.}, however, in the case of the CCQME we require not just the Redfield rates but also their derivatives. Thus, the PT-CCQME retains the numerical simplicity of the standard PT-Redfield equation without requiring high-order dissipators~\cite{Thingna14}.

In summary, the PT-CCQME defined by Eq.~\eqref{eq:CCQME} constitutes our primary result. As a fourth-order quantum master equation in residual interaction~\footnote{Even though the PT-CCQME is fourth-order in the residual interaction strength $\kappa_{mn}$, it inherently captures infinite-order non-perturbative effects of the original frame.}, it generalizes the framework established in Ref.~[\onlinecite{Becker2022}]. The original CCQME expression is readily recovered from this generalized form in the limit of a single-channel interaction, where $S_{mn} = S$ and $V_{mn} = B$.

\section{Spin-Boson Model}
\label{sec:sbm}
The spin-boson model is one of the fundamental paradigms for studying the effects of quantum dissipation, describing a two-level system (spin) coupled to a bosonic environment~\cite{AntoSztrikacs23, Cerisola_2024}. In the original frame, the total Hamiltonian for this model is given by
\begin{align}
\label{eq:sbmHS}
H_S &= \varepsilon \sigma_z + h\sigma_x, \\
\label{eq:sbmHSB}
H_{SB} &= \sigma_z \sum_k g_k(b_k^\dagger + b_k), \\
\label{eq:sbmHB}
H_B &= \sum_k \omega_k b_k^\dagger b_k.
\end{align}
Here, $\sigma_{x,y,z}$ are the Pauli spin-1/2 matrices, and the system and bath interact energetically via the bath's position coordinates.

The spin-boson Hamiltonian can be mapped to the general form introduced in Eqs.~\eqref{eq_bath}--\eqref{eq:system} by restricting the system to two levels ($n=1,2$) and setting $\varepsilon_1=-\varepsilon_2\equiv \varepsilon$, $h_{12}=h_{21}\equiv h$, and $g_{1k}=-g_{2k}\equiv g_k$. Applying the polaron transformation yields the effective Hamiltonians:
\begin{align}
\tilde{H}_S &= \varepsilon \sigma_z + \alpha \sigma_x + \xi\mathbb{I}, \\
\label{eq:sbmPTHSB}
\tilde{H}_{SB} &= h\sigma^{+}V + \mathrm{h.c.}, \\
\tilde{H}_B &= \sum_k \omega_k b_k^\dagger b_k,
\end{align}
where $\mathbb{I}$ is the $2\times2$ identity matrix, $\sigma^+ = (\sigma_x+i\sigma_y)/2$ is the raising operator, and the parameters $\alpha = h\kappa$ and $\xi=-\sum_kg_k^2/\omega_k$. The effective system-bath coupling strength in the polaron frame is 
\begin{equation}
\kappa = \exp\left[-2\sum_k \frac{g_k^2}{\omega_k^2}\coth\left(\frac{\beta\omega_k}{2}\right)\right].
\end{equation}
The residual bath operator $V\equiv V_{12}$, obtained using Eq.~\eqref{eq:V}, is defined as
\begin{equation}
V = \exp\left[2\sum_k \frac{g_k}{\omega_k} (b_k^\dagger -b_k)\right] - \kappa.
\end{equation}
Moreover, $V_{21}\equiv V^\dagger$ and the system operators $S_{12}\equiv\sigma^+$ and $S_{21}\equiv \sigma^-$.

Transitioning to a continuum of bath modes, we introduce the spectral density $J(\omega) = \pi \sum_k g_k^2 \delta(\omega - \omega_k)$, which allows us to express these parameters as integrals:
\begin{align}
\label{eq:kappa}
\kappa &= \exp \left[ -\frac{2}{\pi} \int_{0}^{\infty} d\omega \frac{J(\omega)}{\omega^2} \coth\left(\frac{\beta \omega}{2}\right) \right], \\
\xi &= -\frac{1}{\pi}\int_0^{\infty} d\omega \frac{J(\omega)}{\omega}.
\end{align}
The relevant bath-bath correlation functions satisfy $C_{21}^{12}(\tau)=C_{12}^{21}(\tau)\equiv\langle V(\tau)V^\dagger\rangle = \langle V^\dagger(\tau)V\rangle$ and $C_{12}^{12}(\tau)=C_{21}^{21}(\tau)\equiv\langle V(\tau)V\rangle = \langle V^\dagger(\tau)V^\dagger\rangle$, take the explicit forms:
\begin{align}
\langle V(\tau)V^\dagger\rangle &= \kappa^2\left[e^{-\epsilon(\tau)}-1\right], \\ 
\langle V(\tau)V\rangle &= \kappa^2\left[e^{+\epsilon(\tau)}-1\right],
\end{align}
with the exponent defined as
\begin{align}
\label{eq:epsilon}
\epsilon(\tau) = -\frac{4}{\pi}\int_0^{\infty}d\omega \frac{J(\omega)}{\omega^2}&\left[\coth\left(\frac{\beta\omega}{2}\right)\cos(\omega\tau)\right. \nonumber \\
&\left.- i \sin(\omega\tau)\right].
\end{align}
With the correlation function established, we evaluate their half-sided Fourier transforms denoted as $W_{12}^{21}(\lambda)=W_{21}^{12}(\lambda)\equiv W_{-}(\lambda)$ and $W_{12}^{12}(\lambda)=W_{21}^{21}(\lambda)\equiv W_{+}(\lambda)$. These are explicitly given by,
\begin{align}
    W_{-}(\lambda)&=\kappa^2 \int_0^{\infty}d\tau e^{-i\lambda\tau} \left[e^{-\epsilon(\tau)}-1\right] , \\
    W_{+}(\lambda)&=\kappa^2\int_0^{\infty}d\tau e^{-i\lambda\tau} \left[e^{+\epsilon(\tau)}-1\right].
\end{align}
We can now express the PT-CCQME, derived in Sec.~\ref{sec:PTCCQME}, in the eigenbasis of the effective system Hamiltonian $\tilde{H}_S$. Letting $\ket{\tilde{n}}$ denote the eigenstates of $\tilde{H}_S$ with corresponding eigenenergies $E_n$, we define the energy gaps as $\delta E_{nm} = E_n - E_m$ and the density matrix elements as $\tilde{\rho}_{nm} = \bra{\tilde{n}}\tilde{\rho}_S(t)\ket{\tilde{m}}$. For notational brevity, we drop the time-dependence and the subscript `$S$' from the reduced density matrix ($\tilde{\rho}_S(t) \to \tilde{\rho}$). Furthermore, we assume the Redfield dissipator operates in the Markovian limit ($\tilde{\mathcal{R}}_\infty \to \tilde{\mathcal{R}}$) and omit the `$\mathrm{MFG}$' superscript from the CCQME correction super-operator ($\tilde{Q}^{\mathrm{MFG}}\to \tilde{Q}$). The resulting PT-CCQME reads,
\begin{equation}
    \frac{d\tilde{\rho}_{nm}}{dt} = -i \delta E_{nm}\tilde{\rho}_{nm} + \sum_{u,j=1}^{2}\tilde{\mathcal{R}}_{nm}^{uj}\tilde{\rho}_{uj} - \sum_{u,j=1}^2\tilde{\mathcal{R}}_{nm}^{uj}\sum_{v,l=1}^2\tilde{Q}_{uj}^{vl}\tilde{\rho}_{vl}.
\end{equation}
The Markovian PT-Redfield dissipator [Eq.~\eqref{eq:Redfield}] in the eigenbasis of $\tilde{H}_S$ is given by,
\begin{align}
    \tilde{\mathcal{R}}_{nm}^{uj} &= \sum_{x\neq y=+,-}\sigma^x_{nu}\sigma^y_{jm}\left[W_-(\delta E_{nu})+W_{-}^*(\delta E_{mj})\right] \nonumber \\
    &-\delta_{m,j}\sum_{l=1}^2\sigma^{x}_{nl}\sigma^y_{lu}W_-(\delta E_{lu})-\delta_{n,u}\sum_{l=1}^2\sigma^{x}_{jl}\sigma^y_{lm}W_-^*(\delta E_{lj})\nonumber \\
    &+ \sum_{x=+,-}\sigma^x_{nu}\sigma^x_{jm}\left[W_+(\delta E_{nu})+W_{+}^*(\delta E_{mj})\right] \nonumber \\
    &-\delta_{m,j}\sum_{l=1}^2\sigma^{x}_{nl}\sigma^x_{lu}W_+(\delta E_{lu})-\delta_{n,u}\sum_{l=1}^2\sigma^{x}_{jl}\sigma^x_{lm}W_+^*(\delta E_{lj}). \nonumber
\end{align}
Above $\sigma^{x}_{nm}=\bra{\tilde{n}}\sigma^x\ket{\tilde{m}}$ and $\delta_{u,j}$ is the Kronecker delta. Notably, even though the residual system-bath interaction [Eq.~\eqref{eq:sbmPTHSB}] is expressed in terms of the $\sigma^+$ and $\sigma^-$ system operators, the resulting Redfield dissipator does not inherently reduce to a standard Lindblad form, as it typically would for a pure hopping-type interaction. This deviation arises because the anomalous bath-bath correlations $\langle V(\tau)V\rangle$ and $\langle V^\dagger(\tau)V^\dagger\rangle$ are non-vanishing, which introduces additional fast-rotating terms into the dynamics. 

Finally, the CCQME correction tensor can be compactly written as,
\begin{align}
    \tilde{Q}_{uj}^{vl} &= (1-\delta_{u,j})\frac{\tilde{\mathcal{R}}_{uj}^{vl}}{i\delta E_{uj}} \nonumber \\ 
     & + \delta_{u,j}\delta_{v,l}\sum_{x\neq y = +,-}\left(A_{uv}^{xy,-} - \delta_{u,v}\sum_{r} B_{ru}^{yx,-}\right) \nonumber \\
     & + \delta_{u,j}\delta_{v,l}\sum_{x = +,-}\left(A_{uv}^{xx,+} - \delta_{u,v}\sum_{r} B_{ru}^{xx,+}\right)
\end{align}
where the coefficients $A$ and $B$ encapsulate the derivatives of the half-Fourier transforms and the necessary thermal factors, defined as:
\begin{align}
A_{uv}^{xy,\eta} &= \sigma^x_{uv}\sigma^y_{vu}\mathrm{Im}\left[\frac{\partial W_{\eta}(\delta E_{uv})}{\partial \delta E_{uv}}\right], \\
B_{uv}^{xy,\eta} &= \sigma^y_{vu}\sigma^x_{uv}\left(\mathrm{Im}\left[\frac{\partial W_{\eta}(\delta E_{uv})}{\partial \delta E_{uv}}\right]+\beta W_{\eta}(\delta E_{uv})\right).
\end{align}
\begin{figure}[t!]
    \centering
    \includegraphics[width=\linewidth]{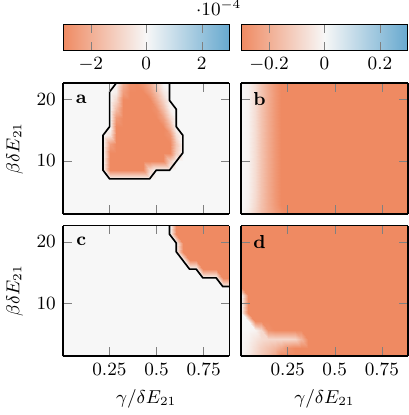}
    \caption{
    Positivity violation captured through the minimum eigenvalue of the reduced density matrix over a large time interval [$t\delta E_{21} \in (0,28)$] for the spin-boson model, plotted as a function of the system-bath coupling strength $\gamma$ and inverse temperature $\beta$. All parameters are expressed in units of the two-level system energy gap $\delta E_{21}$. The system is initialized in the state $\tilde{\rho}_S(0) = (\sigma_z + \mathbb{I})/2$. Negative eigenvalues (red regions) indicate positivity violations. The polaron-transformed equations [PT-Redfield, panel ({\bf a}); and PT-CCQME, panel ({\bf c})] significantly outperform the original-frame Redfield ({\bf b}) and CCQME ({\bf d}) approaches, which consistently violate the positivity of the reduced density matrix. While the PT-Redfield method ({\bf a}) fails at intermediate couplings and low temperatures, the PT-CCQME ({\bf c}) only breaks down at much stronger couplings and lower temperatures, demonstrating its ability to preserve positivity over a substantially broader parameter regime. The system parameter $h=\varepsilon$ with a bath cutoff frequency $\omega_c = \varepsilon$. Black lines in panels ({\bf a}) and ({\bf c}) denote the boundaries of positivity violation.}
    \label{fig:Positivity}
\end{figure}
\begin{figure*}[t!]
    \centering
    \includegraphics[width=\textwidth]{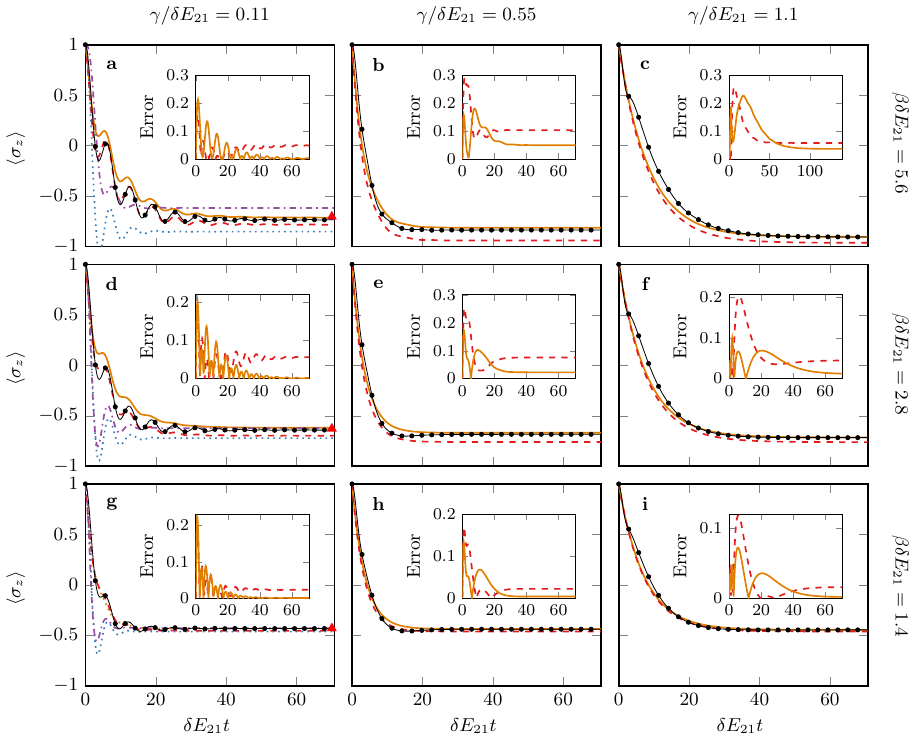}
    \caption{Comparison of the population dynamics $\langle \sigma_z \rangle$ calculated via different master equation approaches against the numerically exact TEMPO method across various system-bath coupling strengths $\gamma$ and inverse temperatures $\beta$. The coupling strength $\gamma$ increases from the weak-coupling (first column) to the strong-coupling (third column) regime, while the inverse temperature $\beta$ decreases from the low-temperature (first row) to the high-temperature (third row) regime. Red solid triangles represent the analytical equilibrium results obtained using the standard Gibbs state [Eq.~\eqref{eq:weakcouplingrho}]. The orange solid lines represent the PT-CCQME results, the red dashed lines represent PT-Redfield, and the black solid lines with circles represent TEMPO. The insets show the error defined as $\mathrm{Error} = |\langle \sigma_z\rangle_{\mathrm{TEMPO}}-\langle \sigma_z\rangle_{\mathrm{X}}|$ with $\mathrm{X}$ being either PT-CCQME (orange solid lines in insets) or PT-Redfield (red dashed lines in insets). Panels ({\bf a}), ({\bf d}), and ({\bf g}) additionally show results for the CCQME (purple dashed dotted lines) and the Redfield (blue dotted lines) equation derived in the original frame. Beyond the weak-coupling regime, both the original-frame CCQME and Redfield equations fail, yielding unphysical values ($|\langle \sigma_z\rangle| > 1$). Details of the TEMPO parameters are given in Appendix \ref{append:tempo}. The TEMPO results are converged by varying the singular value decomposition truncation threshold, time step $dt$, and memory truncation time $t_{\rm mem}$. All other parameters are the same as in Fig.~\ref{fig:Positivity}.}
    \label{fig:Comparison}
\end{figure*}

\section{Numerical Analysis}
\label{sec:numerics}
To concretize the problem, we employ a super-Ohmic spectral density $J(\omega) = \gamma\omega^3\exp[-\omega/\omega_c]$, where $\omega_c$ is the cutoff frequency of the bath. This specific choice ensures that the bath correlation functions are not only analytically tractable but also inherently free of divergences\footnote{Due to the polaron transformation, the integrands for $\kappa$ [Eq.~\eqref{eq:kappa}] and $\epsilon(\tau)$ [Eq.~\eqref{eq:epsilon}] acquire an additional $\omega^{-2}$ factor compared to the quantum master equations in the original frame. For Ohmic and sub-Ohmic spectral densities, this factor introduces divergences at low frequencies, whereas the $\omega^3$ scaling of the super-Ohmic density cleanly regularizes these integrals.}. We note that Ohmic or sub-Ohmic environments can alternatively be accommodated by employing partial polaron transformations~\cite{Teh19, Jang22}. For this super-ohmic spectral density, the functions
\begin{align}
    \kappa &= \exp\left[-\frac{2\gamma}{\pi\beta^2}\left\{-\omega_c^2\beta^2+2\psi_1\left(\frac{1}{\beta\omega_c}\right)\right\}\right], \\
    \xi&=-\frac{2\gamma}{\pi} \omega_c^3, \\
    \epsilon(\tau)&=-\frac{4\gamma}{\pi\beta^2}\left\{\psi_1\left(\frac{1+i\tau\omega_c}{\beta\omega_c}\right)+\psi_1\left(1+\frac{1-i\tau\omega_c}{\beta\omega_c}\right)\right\},
\end{align}
where $\psi_1(z)= \int_0^1\int_0^x dxdy\, x^{z-1}/y(1-x)$ is the trigamma function. Given these expressions, we obtain $W_-(\lambda)$, $W_+(\lambda)$, and their derivatives numerically. For the dynamics, we choose the initial system state to be $\tilde{\rho}_S(0) = (\sigma_z + \mathbb{I})/2$ and focus on the expectation value $\langle \sigma_z\rangle$, an observable that remains invariant under the polaron transformation.

A well-known pathology of the standard Redfield equation and similar perturbative approaches is the violation of complete positivity, a problem that has prompted various corrective strategies \cite{DonatoPRA19,Becker2022,AbbruzzoPRX24}. We first assess whether the PT-CCQME violates positivity and compare its performance against the PT-Redfield equation, as well as the standard Redfield and CCQME approaches in the original frame.

In Fig.~\ref{fig:Positivity}, we plot the smallest eigenvalue of the reduced density matrix, minimized over a long time interval. A negative minimum eigenvalue indicates a violation of positivity. We scan across a broad parameter space, varying the inverse temperature $\beta$ from the deep quantum regime ($\beta\delta E_{21} > 1$) to the high-temperature classical limit ($\beta\delta E_{21} < 1$), alongside varying coupling strengths $\gamma$ from weak- $\gamma/\delta E_{21} < 1$ to strong-coupling $\gamma/\delta E_{21} \approx 1$.

As shown in Fig.~\ref{fig:Positivity}(\textbf{b},\textbf{d}), the original-frame Redfield and CCQME methods consistently violate positivity across the regimes of interest. The PT-Redfield equation shows a marked improvement, though it still breaks down at intermediate coupling strengths within the deep quantum regime. Crucially, the PT-CCQME resolves this intermediate-coupling failure, only exhibiting signs of positivity violation at strong couplings in the deep quantum regime. Because the PT-CCQME incorporates corrections beyond the second-order in the residual coupling, its superior performance over the PT-Redfield equation is expected. The PT-CCQME restricts positivity violations to the extreme parameter regimes where such perturbative master equations are fundamentally expected to fail, providing a much more robust framework than the ad-hoc breakdown seen in the PT-Redfield approach.

Next, we analyze the full dynamics of the system, benchmarking our perturbative equations against numerically exact Time-Evolving Matrix Product Operator (TEMPO) simulations \cite{StrathearnLovett2018} (see Appendix~\ref{append:tempo} for more details on the TEMPO formalism). We choose the initial system state to be $\rho_S(0)=\tilde{\rho}_S(0) = (\sigma_z + \mathbb{I})/2$ and monitor the expectation value $\langle \sigma_z \rangle$. Because $\sigma_z$ commutes with the polaron transformation operator $T$ [Eq. \eqref{eq:transform}], its expectation value is invariant, allowing direct comparison between the polaron and the original frames. The exact TEMPO simulations are performed in the original frame using the total Hamiltonian from Eq. \eqref{eq_total}. Note that TEMPO captures non-Markovian memory effects, which naturally leads to deviations when compared to our Markovian master equations.

The dynamical results are displayed in Fig. \ref{fig:Comparison} across weak (left column), intermediate (center column), and strong (right column) coupling regimes, at low (top row), medium (center row), and high (bottom row) temperatures. The standard Redfield and CCQME approaches formulated in the original frame fail significantly, diverging from the exact TEMPO dynamics even at weak coupling when the temperature is low [Figs.~\ref{fig:Comparison}(\textbf{a}, \textbf{d})]. Furthermore, at intermediate and strong couplings, these original-frame methods yield unphysical steady states ($|\langle\sigma_z\rangle| > 1$) due to positivity violations (omitted from Fig.~\ref{fig:Comparison} for clarity). Consequently, these standard approaches exhibit the least differences from the exact TEMPO in the weak-coupling, high-temperature limit [Fig.~\ref{fig:Comparison}(\textbf{g})].

In contrast, both the PT-Redfield and PT-CCQME frameworks successfully reproduce the exact TEMPO dynamics across the entire coupling spectrum. To quantify this performance, the insets in Fig.~\ref{fig:Comparison} display the absolute dynamical error, defined as $\mathrm{Error} = |\langle \sigma_z\rangle_{\mathrm{TEMPO}}-\langle \sigma_z\rangle_{\mathrm{X}}|$, where X $\in$ {PT-CCQME, PT-Redfield}. While both methods exhibit comparable error magnitudes during the short-time transient dynamics, the PT-CCQME consistently yields smaller errors beyond this initial regime, demonstrating superior accuracy in capturing the long-time relaxation. This advantage is particularly pronounced in the deep quantum regime ($\beta\delta E_{21} = 5.6$) at moderate and strong coupling strengths, where the PT-CCQME distinctly outperforms the PT-Redfield approach. As expected, both polaron-transformed equations perform exceptionally well at high temperatures (bottom row).
\begin{figure}[t!]
    \centering
    \includegraphics[width=\columnwidth]{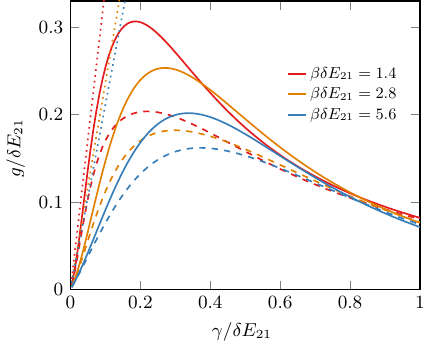}
    \caption{The Liouvillian gap $g$, defined in Eq.~\eqref{eq:gap}, as a function of the system-bath coupling strength $\gamma$ for different inverse temperatures: $\beta \delta E_{21} = 1.4$ (high-temperature regime, red lines), $\beta \delta E_{21} = 2.8$ (moderate-temperature regime, orange lines), and $\beta \delta E_{21} = 5.6$ (deep-quantum regime, blue lines). Solid lines represent the PT-CCQME, while dashed lines correspond to the PT-Redfield equation. We also plot predictions from the original-frame Redfield equation, which deviate significantly from the polaron-transformed approaches (dotted lines). In both the weak-coupling ($\gamma \delta E_{21} \ll 1$) and ultra-strong coupling ($\gamma \delta E_{21} \approx 1$) regimes, the PT-Redfield and PT-CCQME results match, further confirming that the primary discrepancy between these two approaches lies in the intermediate-coupling regime (see discussion surrounding Fig.~\ref{fig:Positivity}). All other parameters are identical to those in Fig.~\ref{fig:Positivity}.}
    \label{fig:thermalization}
\end{figure}

Finally, we study the effect of strong coupling on the thermalization time (also known as the relaxation time) of the spin-boson problem. For Markovian master equations of the form $ d\tilde{\rho}_S(t)/dt = \mathcal{L}[\tilde{\rho}_S(t)]$ studied in this work, we can perform a spectral decomposition to obtain the solution,
\begin{equation}
\tilde{\rho}_S(t) = \sum_j c_j e^{\mu_j t} |\tilde{\Psi}_S^j(t)\rangle.
\end{equation}
Above, we have vectorized the reduced density matrix to transform the Liouvillian $\mathcal{L}$ into a non-Hermitian matrix. Its eigenvalues, $\mu_k$, appear in complex conjugate pairs, with one being exactly zero ($\mu_0 = 0$) \cite{albert:pra14,manzano:av18}. The constants $c_j$ are governed by the overlap of the left-eigenvector of the Liouvillian with the initial condition, and $|\tilde{\Psi}_S^j(t)\rangle$ is the $j$th right-eigenvector of the Liouvillian.

The thermalization time is governed by the inverse of the Liouvillian gap~\cite{Mori20}, which is defined as the negative of the real part of the eigenvalue closest to $\mu_0$, i.e.,
\begin{equation}
\label{eq:gap}
g = - \mathrm{Re}[\mu_1].
\end{equation}
We plot the Liouvillian gap $g$ in Fig.~\ref{fig:thermalization} as a function of the system-bath coupling strength $\gamma$. Even though the PT-Redfield and PT-CCQME yield quantitatively different results, they share a common qualitative feature: at small $\gamma$, increasing the coupling strength increases the Liouvillian gap, indicating that the system will relax faster to its thermal state. However, at large $\gamma$, further increasing the coupling causes the Liouvillian gap to decrease, meaning the system will relax very slowly and exhibit a long-lived metastable state. Because this slowing down of the dynamics is reflected directly in the Liouvillian gap, it is independent of the initial condition and cannot be captured by weak-coupling quantum master equations in the original frame (see dotted lines in Fig.~\ref{fig:thermalization}). This slowdown could potentially be viewed as a Zeno-type effect~\cite{Manzano12, Goyal19}, where a stronger coupling to the environment acts analogously to performing more frequent measurements on the system. It is also worth differentiating this effect from the Mpemba effect in which a system relaxes to the steady-state at different timescales depending on the initial condition~\cite{NavaPRL24, Chirico2026}. The Zeno-type effect studied above does not depend on the initial condition, but our framework leaves open the door to study other such initial-condition-dependent counterintuitive effects (via $c_j$) in the strong system-bath coupling regime.

\section{Concluding Remarks}
\label{sec:discussion}
In this work, we have synergistically combined two powerful techniques to simulate open quantum system dynamics in the strong-coupling regime. Specifically, we applied the polaron transform—which dresses the quantum system with environmental bath modes—to the recently developed Canonically Consistent Quantum Master Equation (CCQME). We benchmarked our newly proposed framework, the PT-CCQME, against the canonical spin-boson model. Our results demonstrate that the PT-CCQME rigorously preserves positivity over a significantly broader range of coupling strengths and temperatures than standard perturbative approaches. Comparisons with numerically exact TEMPO simulations reveal that the PT-CCQME yields accurate dynamics in the deep quantum regime (very low temperatures) and for moderate-to-long times across a broad range of parameters. Furthermore, by construction, the PT-CCQME guarantees that the system relaxes to the correct mean-force Gibbs (MFG) state. We analytically verified that this polaron-transformed MFG state successfully recovers the expected thermodynamic limits in the ultra-weak, ultra-strong, and infinite-temperature regimes. Finally, we demonstrate how our approach can be utilized to estimate thermalization times, revealing a counterintuitive slowing down of relaxation dynamics in the strong system-bath coupling regime.

Despite achieving fourth-order accuracy with respect to the residual system-bath interactions, the PT-CCQME retains the numerical simplicity and efficiency of a standard Markovian master equation. Because it circumvents the high-order super-operators and multidimensional integrals demanded by exact numerical methods such as HEOM or MCTDH, it provides a highly versatile and scalable framework for simulating large many-body quantum systems that were previously computationally prohibitive. While the current formulation relies on a Markovian approximation for the residual interactions in the polaron frame, the underlying theory is highly adaptable. A promising avenue for future research is the development of a fully non-Markovian PT-CCQME, which would systematically extend its validity to structured environments with long bath correlation times. Ultimately, we anticipate that the PT-CCQME will emerge as a practical and robust tool for investigating complex dissipative phenomena across a variety of platforms, ranging from biological light-harvesting complexes to solid-state quantum transport devices.

\begin{acknowledgments}
JT acknowledges fruitful discussions with Alexander Schnell, Tobias Becker, Anton Trushechkin, Andrew Keefe, Carlos A. González-Gutiérrez, and Janet Anders. XX acknowledges the support from the National Natural Science Foundation of China under No. 12305049. DM acknowledges project PID2024-162155OB-I00, funded by MI-CIU/AEI/10.13039/501100011033/ FEDER, UE, as well as the QUANTUM ENIA project call (Quantum Spain project) funded by the Ministry of Economic Affairs and Digital Transformation of the Spanish Government.
\end{acknowledgments}

\section*{Data Availability Statement}
Data is available upon reasonable request from the authors.

\appendix
\section{Derivation of the bath-bath correlation function ($\langle V_{mn}(\tau)V_{m'n'}\rangle$)}
\label{sec:corr2}

The bath-bath correlation function $C_{m'n'}^{mn}= \langle V_{mn}(\tau)V_{m'n'}\rangle$, can be evaluated using the explicit form of $V_{mn}$ from Eq.~\eqref{eq:V}:
\begin{align}
C_{m'n'}^{mn}(\tau)& = \left\langle \left( \Theta_m^\dagger(\tau) \Theta_n(\tau) - \kappa_{mn} \right) \left( \Theta_{m'}^\dagger \Theta_{n'} - \kappa_{m'n'} \right) \right\rangle \nonumber\\
     &= \langle \Theta_m^\dagger(\tau) \Theta_n(\tau) \Theta_{m'}^\dagger \Theta_{n'} \rangle - \kappa_{mn} \langle \Theta_{m'}^\dagger \Theta_{n'} \rangle  \nonumber\\
&- \langle \Theta_m^\dagger(\tau) \Theta_n(\tau) \rangle \kappa_{m'n'} + \kappa_{mn} \kappa_{m'n'}.
\end{align}
Due to the centering of the bath detailed in Sec.~\ref{sec:polaron} $\langle \Theta_m^\dagger(\tau) \Theta_n(\tau) \rangle = \langle \Theta_m^\dagger \Theta_n \rangle = \kappa_{mn}$. Consequently, the correlation function simplifies to
\begin{equation}
    C_{m'n'}^{mn}(\tau)=  \langle \Theta_m^\dagger(\tau) \Theta_n(\tau) \Theta_{m'}^\dagger \Theta_{n'} \rangle - \kappa_{mn} \kappa_{m'n'}.
\end{equation}
Substituting the definition of $\Theta_m$ (see Eq.~\eqref{eq:V}) and using the Heisenberg picture evolution for the free bath operators, $b_k(\tau) = b_k e^{-i\omega_k\tau}$, the first term becomes
\begin{align}
&\langle \Theta_m^\dagger(\tau) \Theta_n(\tau) \Theta_{m'}^\dagger \Theta_{n'} \rangle =\\ 
&\left\langle \exp\left[\sum_k \frac{\delta g_{mn,k}}{\omega_k} \left( b_k^\dagger(\tau) - b_k(\tau) \right)+ \frac{\delta g_{m'n',k}}{\omega_{k}} \left( b_{k}^\dagger - b_{k} \right) \right]\right\rangle. \nonumber
\end{align}

Since the bath is Gaussian, for Gaussian operators $A$ and $B$ we can use the identity $\langle \exp[A+B]\rangle = \exp[\langle A^2\rangle/2 + \langle B^2\rangle/2 + \langle A B\rangle]$ to obtain\footnote{Equivalently, one could also Taylor expand the exponential' and apply Wick's Theorem to evaluate the expectation value.}
\begin{equation}
    \langle \Theta_m^\dagger(\tau) \Theta_n(\tau) \Theta_{m'}^\dagger \Theta_{n'} \rangle= \kappa_{mn} \kappa_{m'n'} \exp[-\epsilon^{m'n'}_{mn}],
\end{equation}
where the exponent is given by
\begin{align}
\epsilon^{m'n'}_{mn} = \sum_k \frac{\delta g_{mn,k} \delta g_{m'n',k}}{\omega_k^2} &\Bigl[ \coth\left(\frac{\beta \omega_k}{2}\right) \cos(\omega_k \tau) \Bigr.\nonumber \\  
&\Bigl.- i \sin(\omega_k \tau) \Bigr]. 
\end{align}
Finally, substituting this back into the expression for $C_{m'n'}^{mn}(\tau)$, we recover the result presented in Eq.~\eqref{eq:C} of the main text:
\begin{equation}
C_{m'n'}^{mn}(\tau) =  \kappa_{mn}\kappa_{m'n'}\left[e^{-\epsilon^{m'n'}_{mn}}-1\right].
\end{equation}

\section{Weak and ultra-strong coupling limits of the polaron transformed mean-force Gibbs state}
\label{sec:WhyPolaron}
In this Appendix, we demonstrate how the mean-force Gibbs (MFG) state, when evaluated using the polaron transformation, successfully recovers the weak-coupling, ultra-strong-coupling, and infinite-temperature (classical) limits~\cite{CresserPRL21, TrushechkinPRA22}.

First, let us explicitly write the MFG state~\footnote{While the MFG state considered here is not strictly normalized, the normalization factor plays no role in the various limits analyzed in this Appendix.} in the polaron basis:
\begin{equation}
    \frac{\mathrm{Tr}_{B}[e^{-\beta \tilde{H}}]}{\tilde{Z}} \simeq \left( \id +  \tilde{Q}^{\mathrm{MFG}} \right) \left[\frac{e^{-\beta \tilde{H}_S}}{\tilde{Z}_S}\right],
\end{equation}
where $\tilde{Q}^{\mathrm{MFG}}$ takes the form given in Eq.~\eqref{eq:Qfinal}.

\subsection*{Ultra-weak coupling limit:}
In this limit, we assume $g_{nk} \rightarrow 0$ for all $n$, where $g_{nk}$ is the system-bath coupling strength in the original frame [see Eq.~\eqref{eq_systembath}]. Consequently, $\delta g_{mn,k} = g_{mk}-g_{nk}$ also approaches zero, yielding
\begin{equation}
 \lim_{g_{uk}\rightarrow 0 \, \forall u} \kappa_{mn} = 1   
\end{equation}
and
\begin{equation}
    \lim_{g_{uk}\rightarrow 0 \, \forall u} C_{m'n'}^{mn}(\tau) = 0.
\end{equation}
Here, we have used Eq.~\eqref{eq:kappa} for $\kappa_{mn}$ and Eq.~\eqref{eq:C} for $C_{m'n'}^{mn}(\tau)$. As $C_{m'n'}^{mn}(\tau) \to 0$ it straightforwardly follows that
\begin{equation}
  \lim_{g_{uk}\rightarrow 0 \, \forall u}\tilde{Q}^{\mathrm{MFG}}[\bullet] = 0.  
\end{equation}
Furthermore, since $\kappa_{mn} \to 1$ in this limit, using Eqs.~\eqref{eq:system} and \eqref{eq:polaronH} yields
\begin{equation}
  \lim_{g_{nu}\rightarrow 0 \, \forall u}\tilde{H}_S = H_{S}.  
\end{equation}
This leads to
\begin{equation}
\label{eq:weakcouplingrho}
  \lim_{g_{uk}\rightarrow 0 \, \forall u}\varrho^{\mathrm{MFG}} = \frac{e^{-\beta H_S}}{Z_S},  
\end{equation}
where $Z_S = \mathrm{Tr}_S[e^{-\beta H_S}]$. As expected, the MFG state ($\varrho^{\mathrm{MFG}}$) correctly reduces to the standard system Gibbs state in the ultra-weak coupling limit.

\subsection*{Infinite temperature limit}
In the high-temperature limit ($\beta\rightarrow 0^+$), applying Eqs.~\eqref{eq:kappa} and~\eqref{eq:C} yields
\begin{align}
    \lim_{\beta \rightarrow 0} \kappa_{mn} &= 0,\nonumber \\
    \lim_{\beta \rightarrow 0} C_{m'n'}^{mn}(\tau) &= 0.
\end{align}
Thus, the effective system Hamiltonian becomes
\begin{equation}
\lim_{\beta\rightarrow 0}\tilde{H}_S = \sum_n\left(\varepsilon_n - \sum_k\frac{g_{nk}^2}{\omega_k}\right)|n\rangle\langle n|.    
\end{equation}
Consequently, the MFG state approaches the maximally mixed state:
\begin{equation}
\lim_{\beta \rightarrow 0}\varrho^{\mathrm{MFG}} = \frac{\id}{N},    
\end{equation}
where $N$ is the dimension of the system's Hilbert space.

\subsection*{Ultra-strong coupling limit}
The ultra-strong coupling limit represents the opposite extreme, where $g_{nk} \rightarrow \infty$ for all $n$. In this regime,
\begin{align}
\lim_{g_{uk} \rightarrow \infty \, \forall u} \kappa_{mn} &= 0, \nonumber \\
\lim_{g_{uk} \rightarrow \infty \, \forall u} C_{m'n'}^{mn}(\tau) &= 0.    
\end{align}
Similar to the infinite-temperature limit, the effective Hamiltonian simplifies to $\lim_{g_{uk} \rightarrow \infty \, \forall u}\tilde{H}_S = \sum_n \varepsilon_n |n\rangle\langle n| - \infty$. Therefore, the ultra-strong coupling MFG state is given by
\begin{equation}
\label{eq:strongcouplingrho}
\lim_{g_{uk} \rightarrow \infty \, \forall u}\varrho^{\mathrm{MFG}} = \frac{e^{-\beta H_S^D}}{Z_S^D},    
\end{equation}
where $H_S^D = \sum_n\varepsilon_n |n\rangle\langle n|$ is the pure dephasing system Hamiltonian and $Z_S^D = \mathrm{Tr}_S[e^{-\beta H_S^D}]$.

This form of the ultra-strong coupling MFG state was previously obtained in Ref.~[\onlinecite{CresserPRL21}] using a completely different methodology; however, the polaron transformation provides a remarkably simple alternative derivation. Physically, this indicates that the ultra-strong MFG state remains a Gibbs state, but with respect to the projected effective Hamiltonian $H_S^D$. Because the system-bath interaction is overwhelmingly strong, the system operator present in the interaction dominates, effectively projecting the Hamiltonian into the $\{|n\rangle\}$ basis and yielding $H_S^D$.

Surprisingly, the polaron transformation successfully recovers the correct MFG state in the extreme limits of both weak and ultra-strong coupling—a feat that is impossible to achieve when working strictly in the original frame as also shown in Ref.~[\onlinecite{TrushechkinPRA22}]. This robust analytical property also implies that the PT-CCQME correctly relaxes to the appropriate steady state in these limits, as the second-order polaron transformed MFG state inherently serves as the steady state of the PT-CCQME. Consequently, this demonstrates the reliability of the PT-CCQME technique, suggesting it is well-suited for weak- and ultra-strong coupling strengths, and providing a powerful tool for capturing strong system-bath coupling effects in interacting quantum many-body systems.

\section{Derivation of the mean-force Gibbs state $\tilde{Q}^{\mathrm{MFG}}$ induced correction}
\label{sec:Q}

In this Appendix, we rederive $\tilde{Q}^{\mathrm{MFG}}$ for the system-bath interaction given by Eq.~\eqref{eq:systembathint}. We begin by expanding $e^{-\beta \tilde{H}}$ up to second order in the residual interaction $\tilde{H}_{SB}$:
\begin{align}
    e^{-\beta \tilde{H}} &\simeq e^{-\beta (\tilde{H}_S+\tilde{H}_B)}\left[\id + \int_{0}^{\beta} d\mu \tilde{H}_{SB}(-i\mu)\right. \nonumber \\
    & + \left.\int_{0}^{\beta} d\mu\int_0^\mu d\xi\tilde{H}_{SB}(-i\mu)\tilde{H}_{SB}(-i\xi)\right].
\end{align}
Above, $\tilde{H}_{SB}(-i\mu)$ is the imaginary-time interaction-picture system-bath interaction Hamiltonian and we have used the Kubo identity to expand the exponential (obtained by iterating $\exp[\alpha(\hat{A}+\hat{B})] = \exp[\alpha \hat{A}] [\mathbb{I}+\int_0^{\alpha}d\alpha' \exp[-\alpha'\hat{A}] \hat{B} \exp[\alpha'(\hat{A}+\hat{B})]$ to expand in powers of $\hat{B}$).

Using the form of $\tilde{H}_{SB}$ given in Eq.~\eqref{eq:systembathint}, tracing over the bath degrees of freedom, and using Eq.~\eqref{eq:MFG} gives,
\begin{align}
\tilde{Q}^{\mathrm{MFG}} \left[ \frac{e^{-\beta \tilde{H}_S}}{\tilde{Z}_S} \right] &= \frac{e^{-\beta \tilde{H}_S}}{\tilde{Z}_S} \int_{0}^{\beta} d\mu \int_{0}^{\mu} d\xi \\
&\sum_{\substack{n \neq m \\ n' \neq m'}} S_{mn}(-i\mu) S_{m'n'}(-i\xi) C^{mn}_{m'n'}(-i(\mu-\xi)). \nonumber
\end{align}
Changing integration variables to $s=(\mu+\xi)/2$ and $v=\mu-\xi$ yields,
\begin{widetext}
\begin{equation}
\tilde{Q}^{\mathrm{MFG}} \left[ \frac{e^{-\beta \tilde{H}_S}}{\tilde{Z}_S} \right] = \frac{e^{-\beta \tilde{H}_S}}{\tilde{Z}_S} \int_{0}^{\beta} dv \int_{v/2}^{\beta-v/2} ds \sum_{\substack{n \neq m \\ n' \neq m'}} e^{s\tilde{H}_S}S_{mn}\left(-\frac{iv}{2}\right) S_{m'n'}\left(\frac{iv}{2}\right)e^{-s\tilde{H}_S} C^{mn}_{m'n'}(-iv).
\end{equation}
\end{widetext}

We now introduce the projection operators defined below Eq.~\eqref{eq:Qfinal}: $\mathbf{P} = \sum_{n\neq m}\mathbf{P}_{nm}$ with $\mathbf{P}_{nm}\bullet = |\tilde{n}\rangle\langle \tilde{n}| \bullet |\tilde{m}\rangle\langle \tilde{m}|$ and $\mathbf{P^c}=\sum_n \mathbf{P}_{nn}$. Furthermore, noting the identities 
\begin{align}
    \mathbf{P}\int ds e^{s\tilde{H}_S} \bullet e^{-s\tilde{H}_S} & = \mathbf{P}\frac{1}{\tilde{\Delta}} e^{s\tilde{H}_S}\bullet e^{-s\tilde{H}_S}, \\
    \mathbf{P^c}\int ds e^{s\tilde{H}_S} \bullet e^{-s\tilde{H}_S} & = \mathbf{P^c} s, 
\end{align}
we project $\tilde{Q}^{\mathrm{MFG}}$ into its coherent and complimentary subspaces to give,
\begin{widetext}
\begin{align}
\label{eq:Qimagfin}
    \tilde{Q}^{\mathrm{MFG}}\left[ \frac{e^{-\beta \tilde{H}_S}}{\tilde{Z}_S} \right] &= \mathbf{P}\frac{1}{\tilde{\Delta}} \sum_{\substack{n \neq m \\ n' \neq m'}} \int_0^{\beta} d\nu \left[S_{mn}S_{m'n'}(i\nu)\frac{e^{-\beta \tilde{H}_S}}{\tilde{Z}_S} - \frac{e^{-\beta \tilde{H}_S}}{\tilde{Z}_S} S_{mn}(-i\nu)S_{m'n'} \right]C_{m'n'}^{mn}(-i\nu) \nonumber \\
    &+ \mathbf{P^c} \frac{e^{-\beta \tilde{H}_S}}{\tilde{Z}_S} \sum_{\substack{n \neq m \\ n' \neq m'}} \int_0^{\beta}d\nu S_{mn} S_{m'n'}(i\nu) (\beta - \nu) C_{m'n'}^{mn}(-i\nu).
\end{align}
\end{widetext}

Expressing the above equation in the eigenbasis of $\tilde{H}_S$ and using the relation between the integrals of the imaginary- and real-time correlation functions [Eq.~\eqref{eq:relationCimCre}] derived in Appendix~\ref{append:correlationrelation}, namely,
\begin{align}
\label{eq:relation}
    - \int_0^{\beta} du\, C^{mn}_{m'n'}(-i u) e^{-\lambda u}&=  \mathrm{Im} \left[ W^{mn}_{m'n'}(\lambda) \right] \nonumber \\
    &+ e^{-\beta \lambda} \mathrm{Im}\left[ W_{mn}^{m'n'}(-\lambda) \right]
\end{align}
with  $W^{mn}_{m'n'}(\lambda) = \int_0^{\infty} d\tau C^{mn}_{m'n'}(\tau) e^{-i\lambda\tau}$. Rearranging the terms, we arrive at the final expression for $\tilde{Q}^{\mathrm{MFG}}$,
\begin{equation}
\label{eq:QfinAppend}
\tilde{Q}^{\mathrm{MFG}}\left[ \frac{e^{-\beta \tilde{H}_S}}{\tilde{Z}_S} \right] = \mathbf{P}\frac{1}{i\tilde{\Delta}}\left[\tilde{\mathcal{R}}_{\infty}\left[ \frac{e^{-\beta \tilde{H}_S}}{\tilde{Z}_S} \right]\right] + \tilde{\mathcal{L}}\left[\mathbf{P^c}\, \frac{e^{-\beta \tilde{H}_S}}{\tilde{Z}_S} \right].
\end{equation} 
The super-operator $(i\tilde{\Delta})^{-1}[\bullet]$ is defined such that
\begin{equation}
    \mathbf{P} \frac{1}{i\tilde{\Delta}}[\bullet] := \sum_{n \neq m} \frac{1}{i\delta E_{nm}}|\tilde{n}\rangle\langle \tilde{n}| \bullet |\tilde{m}\rangle\langle \tilde{m}|.
\end{equation}
Above, $\tilde{\mathcal{R}}_\infty$ is the Redfield dissipator. In transitioning to the final expression, we utilized the fact that the real parts of $W_{m'n'}^{mn}(\lambda)$ satisfy local detailed balance, as proven in Appendix~\ref{sec:LDB}. Consequently, the Redfield dissipator containing only the real part of $W_{m'n'}^{mn}(\lambda)$ vanishes when acting on the Gibbs state $e^{-\beta \tilde{H}_S}/\tilde{Z}_S$. This allows us to construct the full Redfield dissipator $\tilde{\mathcal{R}}_{\infty}$, even though only the imaginary part of $W_{m'n'}^{mn}(\lambda)$ explicitly appears when substituting the correlation relation. The generalized Lindblad dissipator $\tilde{\mathcal{L}}$ takes the form,
\begin{equation}
\tilde{\mathcal{L}}[\bullet] = \sum_{u,j} \Gamma_{uj}L_{uj} \bullet L_{uj}^{\dagger} - \frac{\Gamma_{uj}'}{2}\{ L_{uj}^\dagger L_{uj},\bullet \} ,
\end{equation}
with $L_{uj}=|\tilde{u}\rangle\langle \tilde{j}|$ being the jump operators and the generalized rates 
\begin{align}
\Gamma_{uj} &= \sum_{\substack{n \neq m \\ n' \neq m'}} \langle \tilde{u}|S_{mn}|\tilde{j}\rangle\langle \tilde{j} |S_{m'n'}|\tilde{u}\rangle \mathrm{Im}\left[\frac{\partial W^{m'n'}_{mn}(\delta E_{uj})}{\partial \delta E_{uj}}\right], \nonumber \\
\Gamma_{uj}' &= \sum_{\substack{n \neq m \\ n' \neq m'}} \langle \tilde{u}|S_{m'n'}|\tilde{j}\rangle\langle \tilde{j} |S_{mn}|\tilde{u}\rangle \mathrm{Im}\left[\frac{\partial W^{mn}_{m'n'}(\delta E_{uj})}{\partial \delta E_{uj}} \right. \nonumber\\
&\left.\qquad\qquad\qquad\qquad\qquad\qquad +\beta W^{mn}_{m'n'}(\delta E_{uj})\right].
\end{align}
We note that this formulation of the mean-force Gibbs state induced correction $Q^{\textrm {MFG}}$ differs slightly from the original version presented in Ref.~[\onlinecite{Becker2022}]. In the original framework, the rates are symmetric ($\Gamma'_{uj} = \Gamma_{uj}$), necessitating an additional term involving the energy derivative of the density matrix. Our current variant mathematically simplifies this structure by absorbing that complexity into two distinct rates, $\Gamma'_{uj}$ and $\Gamma_{uj}$. The dynamical differences between these two approaches are minor when compared against exact TEMPO simulations for the spin-boson model discussed in the main text.

\section{Relation between imaginary- and real-time correlation functions}
\label{append:correlationrelation}
A critical step in deriving the PT-CCQME is establishing the relationship between the integrals of the imaginary-time and real-time correlation functions. While this relation can generally be obtained via a Wick rotation, we explicitly derive this relation for the Polaron framework described in this manuscript.

We begin with the integral of the imaginary-time correlation function $C_{m'n'}^{mn}(-iu)$, which can be separated as follows:
\begin{align}
    \int_{0}^{\beta} du C_{m'n'}^{mn}(-iu) e^{-u\lambda} &= \int_{0}^{\infty}du C_{m'n'}^{mn}(-iu) e^{-u\lambda} \nonumber \\
    & -\int_{\beta}^{\infty}du C_{m'n'}^{mn}(-iu) e^{-u\lambda}.
\end{align}
Changing the integration variable to $v=u-\beta$ in the second integral on the right-hand side yields,
\begin{align}
\label{eq:Cim}
    \int_{0}^{\beta} du C_{m'n'}^{mn}(-iu) e^{-u\lambda} &= \int_{0}^{\infty}du C_{m'n'}^{mn}(-iu) e^{-u\lambda} \nonumber \\
    & -e^{-\beta \lambda}\int_{0}^{\infty}dv C_{m'n'}^{mn}(-iv-i\beta) e^{-v\lambda}.
\end{align}
Next, we use the explicit form of the bath-bath correlation function from Eq.~\eqref{eq:C} to give,
\begin{equation}
C_{m'n'}^{mn}(\tau) =  \kappa_{mn}\kappa_{m'n'}\left[e^{-\epsilon^{m'n'}_{mn}}-1\right],
\end{equation}
where the exponent $\epsilon_{m'n'}^{mn}$ is defined in Eq.~\eqref{eq:epsilon} as
\begin{equation}
\epsilon_{m'n'}^{mn}= \int_{0}^{\infty} \frac{d\omega}{\pi} \frac{J(\omega)}{\omega^2} \left[ n_\omega e^{i\omega\tau}
+ (n_\omega +1) e^{-i\omega\tau} \right].
\end{equation}
Here $n_{\omega} = [\exp[\beta\omega]-1]^{-1}$ is the Bose-Einstein distribution, and we have assumed a continuous spectral density $J(\omega) = \pi \sum_k \delta g^2_{mn,k}  \, \delta(\omega - \omega_k)$. 

By substituting this form of the correlation function, Taylor expanding the exponential, exchanging the order of the time ($u$ or $v$) and frequency ($\omega$) integrations, and explicitly evaluating the time integrals, we obtain the final result:
\begin{align}
\label{eq:relationCimCre}
    - \int_{0}^{\beta} du\, C^{mn}_{m'n'}(-i u) e^{-\lambda u} &= \mathrm{Im} \left[ \int_{0}^{\infty} d\tau\, C^{mn}_{m'n'}(\tau) e^{-i \lambda \tau} \right] \nonumber \\
    &+ e^{-\beta \lambda} \mathrm{Im} \left[ \int_{0}^{\infty} d\tau\, C_{mn}^{m'n'}(\tau) e^{i \lambda \tau} \right].
\end{align}
The equation above relates the integrals of the imaginary-time bath-bath correlation function on the left-hand side to the Fourier-Laplace transforms of the real-time correlation functions on the right-hand side. This relationship is critical, as it ensures that $\tilde{Q}^{\mathrm{MFG}}$ can be evaluated using the exact same half-sided Fourier transforms, $W_{m'n'}^{mn}(\lambda) = \int_0^{\infty}d\tau\, C_{m'n'}^{mn}(\tau) e^{-i\lambda \tau}$, that are already utilized in the standard PT-Redfield framework. Ultimately, connecting the imaginary-time and real-time integrals of the bath correlation functions guarantees that solving the PT-CCQME introduces no additional computational complexity compared to the standard PT-Redfield equation.

\section{Proving Local-Detailed Balance}
\label{sec:LDB}
In order to prove local-detailed balance we begin with the form of the $W_{m'n'}^{mn}(\lambda)$ given below Eq.~\eqref{eq:Lindbladrates}, i.e., 
\begin{equation}
    W^{mn}_{m'n'}(\lambda) = \int_0^{\infty} d\tau\, C^{mn}_{m'n'}(\tau) e^{-i\lambda\tau},
\end{equation}
with
\begin{equation}
C_{m'n'}^{mn}(\tau) =  \kappa_{mn}\kappa_{m'n'}\left[e^{-\epsilon^{m'n'}_{mn}}-1\right].
\end{equation}
Local detailed-balance is defined as 
\begin{equation}
    \frac{\mathrm{Re}[W^{mn}_{m'n'}(\lambda)]}{\mathrm{Re}[W^{mn}_{m'n'}(-\lambda)]} = e^{-\beta \lambda}.
\end{equation}
For a continuous $J(\omega) = \pi \sum_{k}\delta g_{mn,k}^2\delta (\omega-\omega_k)$ and odd spectral density $J(-\omega)=-J(\omega)$ the exponent $\epsilon_{mn}^{m'n'}$ given in Eq.~\eqref{eq:epsilon} can be massaged into
\begin{equation}
\epsilon^{m'n'}_{mn} = \int_{-\infty}^{\infty} \frac{d\omega}{\pi} \frac{J(\omega)}{\omega^2} n_{\omega}e^{i\omega\tau},
\end{equation}
where $n_{\omega} = [\exp[\beta\omega]-1]^{-1}$ is the Bose-Einstein distribution. Thus, the function
\begin{align}
    W^{mn}_{m'n'}(\lambda) =  \kappa_{mn}\kappa_{m'n'}   &\int_0^{\infty} d\tau\, e^{-i\lambda\tau} \\
    &\left(\exp\left[-\int_{-\infty}^{\infty} \frac{d\omega}{\pi} \frac{J(\omega)}{\omega^2} n_\omega e^{i\omega\tau}\right]-1\right).\nonumber 
\end{align}
In order to prove local-detailed balance our goal is to swap the $\tau$ and $\omega$ integrals. However, this is not straightforward since the $\omega$ integral is in the exponent. Hence, we begin by defining a function
\begin{equation}
    A(\omega) = -\frac{1}{\pi} \frac{J(\omega)}{\omega^2} n_\omega.
\end{equation}
Taylor expanding the exponential containing the $\omega$ integral yields,
\begin{widetext}
\begin{equation}
    W_{m'n'}^{mn}(\lambda) = \kappa_{mn}\kappa_{m'n'} \int_{0}^{\infty}d\tau e^{-i\lambda\tau}\Biggl[\int_{-\infty}^{\infty}d\omega A(\omega)e^{i\omega\tau}
    + \frac{1}{2!}\int_{-\infty}^{\infty}d\omega A(\omega)e^{i\omega\tau}\int_{-\infty}^{\infty} d\omega'A(\omega')e^{i\omega'\tau}+ \cdots \Biggr].
\end{equation}    
\end{widetext}
Now since the $\omega$ integral is not in the exponent we can swap the $\omega$ and $\tau$ integrals and use the Sokhotski-Plemelj formula~\footnote{The Sokhotski-Plemelj reads $\int_{0}^\infty d\tau e^{\pm i \Omega \tau} = \delta(\Omega)\pm i \mathcal{P}\frac{1}{\Omega}$, where $\mathcal{P}$ represents the principal value part.} to obtain the real-part of $W_{m'n'}^{mn}(\lambda)$ as,
\begin{align}
\label{eq:Wnum}
    \mathrm{Re}[ W_{m'n'}^{mn}(\lambda)] &= \kappa_{mn}\kappa_{m'n'} \Biggl[ A(\lambda) \Biggr.\\
    &\Biggl.+ \frac{1}{2!}\int_{-\infty}^{\infty}d\omega A(\omega)A(\lambda-\omega)+ \cdots \Biggr].\nonumber 
\end{align}
Next, multiplying both sides above by $\exp[\beta \lambda]$ and noting that $n_x \exp[\beta x] = n_x + 1$ and $n_{-x} = -(n_x+1)$ we obtain,
\begin{align}
\label{eq:Wden}
     \mathrm{Re}[ W_{m'n'}^{mn}(\lambda)]e^{\beta \lambda} &= \kappa_{mn}\kappa_{m'n'} \Biggl[ A(-\lambda) \Biggr. \\
     & \Biggl.+ \frac{1}{2!}\int_{-\infty}^{\infty}d\omega A(\omega)A(-\lambda-\omega)+ \cdots \Biggr]. \nonumber
\end{align}
Thus, using Eqs.~\eqref{eq:Wnum} and~\eqref{eq:Wden} gives us the local-detailed balance condition, namely,
\begin{equation}
    \mathrm{Re}[ W_{m'n'}^{mn}(\lambda)]e^{\beta \lambda} = \mathrm{Re}[ W_{m'n'}^{mn}(-\lambda)]. 
\end{equation}

\section{Brief introduction to time-evolving matrix product operator method \label{append:tempo}}

The time-evolving matrix product operator (TEMPO) method is a numerically exact approach to simulate the dynamics of a non-Markovian open quantum system~\cite{StrathearnLovett2018}. In this work, we benchmark the results from various master equations with the TEMPO method. Its core idea is to discretize the Feynman-Vernon influence functional with the quasi-adiabatic propagator path integral (QuAPI) scheme \cite{MakriMakarov1995, MakriMakarov1995a}, and subsequently to make it a tensor network in the temporal domain. This tensor-network representation allows simulation of dynamics over much longer memory times compared to standard QuAPI, due to a controllable bond dimension in the influence functional~\cite{VilkoviskiyPRB24}. Moreover, unlike methods such as the HEOM, it offers flexibility in dealing with various spectral functions. In the following, we give a short introduction to the TEMPO method, focusing on the system-bath interaction $H_{SB}=S\otimes B$ (see Eq.~\eqref{eq:sbmHSB} for the $H_{SB}$ of the spin-boson model).

The corresponding reduced density matrix of the original frame of the system $\rho_S(t)$ can be expressed as a real-time path integral on the Keldysh contour. By introducing the forward and backward path variables $\mathbf{s} \equiv (s^+,s^-)$ the reduced density matrix at time $t$ is given by 
\begin{align}
    \rho_S(t) = Z_{B}\int \mathcal{D}[\mathbf{s}]K[\mathbf{s}]I[\mathbf{s}],
\end{align}
where $Z_B$ is the partition function of the bath. Above, $K[\mathbf{s}]$ is the system propagator and $I[\mathbf{s}]$ is the influence functional. The explicit form of the influence functional for the bosonic environment can be found in Refs. [\onlinecite{MakriMakarov1995}] and [\onlinecite{MakriMakarov1995a}].

The path integral is then discretized using the QuAPI scheme with time step $\delta t$ such that $t = N \delta t$ and $\mathbf{s}(k \delta t) \equiv \mathbf{s}_k = (s_k^+, s_k^-)$. The reduced density matrix then becomes
\begin{align}
    \rho_S(\mathbf{s}_N) = \sum_{\mathbf{s}_0, \cdots, \mathbf{s}_{N-1}} K^{\mathbf{s}_0,\cdots, \mathbf{s}_N}\,I^{\mathbf{s}_0,\cdots, \mathbf{s}_N},
\end{align}

The free-system propagator factorizes locally in time, namely,
\begin{align}
  K^{\mathbf{s}_0,\cdots, \mathbf{s}_N} =   \tilde{K}^{\mathbf{s}_0,\mathbf{s}_1} \,\cdots \,   \tilde{K}^{\mathbf{s}_{N-1},\mathbf{s}_N}.
\end{align}
where $\tilde{K}^{\mathbf{s}_{k-1},\mathbf{s}_k} = \langle s^+_k \rvert e^{-i H_{S} \delta t} \lvert s^-_{k-1} \rangle  \langle s^+_{k-1} \rvert e^{i H_{S} \delta t} \lvert s^-_{k} \rangle$, and the influence functional can be reorganized into factors coupling time slices separated by $\Delta k$ as~\cite{MakriMakarov1995}
\begin{align}
    I^{\mathbf{s}_0,\cdots, \mathbf{s}_N} =
   \prod^N_{\Delta k=0} \prod^{N-\Delta k}_{k=0} I_{\Delta k}^{\mathbf{s}_{k},\mathbf{s}_{k+\Delta k}}.
\end{align}
This structure motivates a recursive propagation of augmented density tensor, $F^{\mathbf{s}_0,\cdots, \mathbf{s}_N}
:= K^{\mathbf{s}_0,\cdots, \mathbf{s}_N} \, I^{\mathbf{s}_0,\cdots, \mathbf{s}_N}$, such that
\begin{align}
F^{\mathbf{s}_0,\cdots, \mathbf{s}_{k+1}} =
F^{\mathbf{s}_0,\cdots, \mathbf{s}_{k}}\,
A^{\mathbf{s}_0,\cdots, \mathbf{s}_{k+1}}.
\end{align}
Above $A^{\mathbf{s}_0,\cdots, \mathbf{s}_{k+1}}$ collects the short-time propagator between $k$ and $k+1$ and all influence factors involving the newest index $k+1$, i.e.,
\begin{align}
A^{\mathbf{s}_0,\cdots, \mathbf{s}_{k+1}}
= \tilde{K}^{\mathbf{s}_k,\mathbf{s}_{k+1}}
\prod_{\ell=0}^{k+1}
I_{k+1-\ell}^{\mathbf{s}_\ell, \mathbf{s}_{k+1}}.
\end{align}
Starting from $F^{\mathbf{s}_0}=\rho_S(\mathbf{s}_0)\,I_0^{\mathbf{s}_0,\mathbf{s}_0}$ and iterating the above recursion yields $F^{\mathbf{s}_0,\cdots,\mathbf{s}_N}$, and hence $\rho_S(\mathbf{s}_N)$ by summation over the intermediate indices.

To express the recursion as a tensor-network update, it is convenient to introduce an MPO by appropriately inserting Kronecker deltas and defining
$B^{\mathbf{s}_0,\cdots,\mathbf{s}_{k+1}}_{\mathbf{r}_0,\cdots,\mathbf{r}_k}
:=
A^{\mathbf{s}_0,\cdots,\mathbf{s}_{k+1}}
\prod_{j=0}^{k}\delta^{\mathbf{s}_j}_{\mathbf{r}_j}$, such that the update takes the compact form
\begin{align}
F^{\mathbf{s}_0,\cdots,\mathbf{s}_{k+1}}
=
\sum_{\mathbf{r}_0,\cdots,\mathbf{r}_k}
F^{\mathbf{r}_0,\cdots,\mathbf{r}_k}\,
B^{\mathbf{s}_0,\cdots,\mathbf{s}_{k+1}}_{\mathbf{r}_0,\cdots,\mathbf{r}_k}.
\end{align}
A direct implementation is exponentially costly because $F^{\mathbf{s}_0,\cdots,\mathbf{s}_k}$ has rank $k+1$, leading to scaling $\sim M^{2(k+1)}$ with the system Hilbert-space dimension $M$. TEMPO circumvents this by representing the path tensor as a matrix product state (MPS),
\begin{align}
F^{\mathbf{s}_0,\cdots,\mathbf{s}_k}
=
\sum_{\{\alpha\}}
F^{\mathbf{s}_0}_{\alpha_0}
F^{\mathbf{s}_1}_{\alpha_0\alpha_1}
\cdots
F^{\mathbf{s}_{k-1}}_{\alpha_{k-2}\alpha_{k-1}}
F^{\mathbf{s}_k}_{\alpha_{k-1}},
\end{align}
and expressing the update $B$ as an MPO. The time evolution is then implemented by repeated MPO--MPS multiplication, with controlled bond-dimension truncation. In practice, for baths with finite correlation time, influence factors beyond a memory cutoff time $t_{mem}$ can be neglected, yielding efficient and accurate simulations at long times. In Fig.~\ref{fig:Comparison}, the results for TEMPO are computed with time step $dt=0.1$, singular value decomposition truncation threshold $\varepsilon_{\rm svd} = 10^{-8}$, and memory truncation time $\delta E_{21} t_{\rm mem}=25\sqrt{2}$. 
\nocite{*}
\bibliographystyle{aipnum4-1}
\bibliography{biblio}
\end{document}